%Paper: gr-qc/9309003
%From: Jorge Pullin <pullin@phys.psu.edu>
%Date: Wed, 1 Sep 93 23:22:13 EDT

\magnification=\magstep1
\vbadness=10000
\parskip=\baselineskip
\parindent=10pt
\centerline{\bf MATTERS OF GRAVITY}
\bigskip
\bigskip
\line{Number 2 \hfill Fall 1993}
\bigskip
\bigskip
\bigskip
\centerline{\bf Table of Contents}
\bigskip
\hbox to 6truein{Editorial {\dotfill} 1}
\hbox to 6truein{Correspondents {\dotfill} 3}
\hbox to 6truein{Some recent work in general relativistic
astrophysics  {\dotfill} 4}
\hbox to 6truein{Two dimensional black holes {\dotfill} 6}
\hbox to 6truein{Resonant-mass gravitational wave
detectors: an update {\dotfill} 8}
\hbox to 6truein{Universality and scaling in gravitational collapse
 {\dotfill} 9}
\hbox to 6truein{Gravitational Wave memories upgraded {\dotfill} 11}
\hbox to 6truein{Conference report: quantum aspects of
black holes{\dotfill} 13}
\hbox to 6truein{Conference report: knots and quantum gravity{\dotfill} 14}
\hbox to 6truein{Conference report: deterministic chaos in GR{\dotfill} 17}
\bigskip
\bigskip
\bigskip
\centerline{\bf Editorial}
\medskip
\centerline{Jorge Pullin}
\centerline{Center for Gravitational Physics and Geometry}
\centerline{Penn State University}
\bigskip

        Welcome to the second issue of {\it Matters of Gravity}, a
newsletter for the gravitational physics community of the United
States.

	Two years have passed since the appearance of the first issue.
In the meantime, Peter Saulson has stepped down as editor and I took
over recently to try to restore continuity to the Newsletter. I would
try to assure that we get published at least twice a year, once in the
fall and in the spring. Publication dates are September and February
1st.

	As before, the intention of the newsletter is to try to bring
a sense of community to the people working in gravitational physics in
the US. Everyone is welcome to contribute about any topic dealing with
gravitaional physics. Articles are not to exceed two pages in length
and should avoid talking about one's own work. In this issue we
incorporate a new format of articles: reports about conferences.
Conference organizers and anyone else interested in writing such
reports are welcome to contribute.

	Putting together a newsletter like this implies a certain non
negligible amount of effort. It would be healthier for the newsletter
if we had more than one editor. This would also help strike some
balance in the topics covered. It would be extremely helpful if we had
an editor with experimental interests. Anyone willing to invest some
effort is welcome to join in as editor.

	As before, we keep a group of correspondents to try to get
input from as wide an audience as possible. Many thanks to them for
their cooperation in putting this issue together. As in the previous
issue, we invite anyone who wants to serve as correspondant to join
in.

        This issue has been distributed in two forms. A paper edition,
produced in \TeX, was mailed to a list we derived mainly from the
NSF's list of sponsored workers in gravitational physics. A larger
number of you are receiving {\it Matters of Gravity} via e-mail. Due
to technical problems in PSU we sent the Newsletter out on Malcolm
MacCallum's list and posted it in the LANL gr-qc bulletin board. Since
the Newsletter is focused on the US community, we should plan to have
a more focused distribution mechanism for the future. The newsletter
is also available as a postscript file for those interested.

	I would like to end by thanking the contributors. Writing
these short articles requires more effort than people may think.

        Please also send me your comments on this second issue of the
newsletter, as well as suggestions for articles.

\medskip
\leftline{Jorge Pullin}
\smallskip
\leftline{Center for Gravitational Physics and Geometry}
\leftline{The Pennsylvania State University}
\leftline{University Park, PA 16802-6300}
\smallskip
\leftline{Fax: (814)863-9608}
\leftline{Phone (814)863-9597}
\leftline{Internet: pullin@phys.psu.edu}

\vfill
\eject
\centerline{\bf Correspondents}
\medskip

\item{1.} John Friedman and Kip Thorne: Relativistic Astrophysics,
\item{2.} Jim Hartle: Quantum Cosmology and Related Topoics
\item{3.} Gary Horowitz: Interface with Mathematical High Energy Physics,
    including String Theory
\item{4.} Richard Isaacson: News from NSF
\item{5.} Richard Matzner: Numerical Relativity
\item{6.} Abhay Ashtekar and Ted Newman: Mathematical Relativity
\item{7.} Bernie Schutz: News From Europe
\item{8.} Lee Smolin: Quantum Gravity
\item{9.} Cliff Will: Confrontation of Theory with Experiment
\item{10.} Peter Bender: Space Experiments
\item{11.} Riley Newman: Laboratory Experiments
\item{12.} Peter Michelson: Resonant Mass Gravitational Wave Detectors
\item{13.} Robbie Vogt: LIGO Project

\vfill
\eject

\centerline{\bf Some recent work in general relativistic astrophysics}
\medskip
\centerline{John Friedman, University of Wisconsin - Milwaukee}
\bigskip
\bigskip

Stephen Detweiler has been studying the evolution of a binary black
hole system by examining solutions of the Einstein equations which are
periodic when viewed from some rotating frame of reference.   The
approach generalizes an earlier study with Blackburn of solutions
stationary in a rotating frame.  When the fractional energy loss per
period is small, the radiation field near the binary system is small;
and the Detweiler-Blackburn solutions with equal amounts of ingoing and
outgoing radiation can approximate the evolution of a binary black hole
system from the time when the holes are far apart, through the stage of
slow evolution caused by gravitational radiation reaction, up until the
time when the radiation reaction timescale is comparable to the
dynamical timescale.

Detweiler's approximation allows one to use a variational principle to
estimate key features of a binary system.  Accurate estimates are
claimed for the relationship between the total energy and angular
momentum of the system, the masses and angular momenta of the
components, the rotational frequency of the frame of reference in which
the system is periodic, the frequency of the periodicity of the system,
the stationary mass and current moments and the amplitude and phase of
each multipole component of gravitational radiation.

*

The study of nonradial oscillations of relativistic stars, begun 25
years ago by Thorne and his colleagues, has recently been renewed in
work by Chandrasekhar and Ferrari (For a review, see V. Ferrari,
``Non-radial oscillations of stars in general relativity: a scattering
problem,'' in Classical General Relativity, ed. S. Chandrasekhar,
Oxford, 1993).  Chandrasekar and Ferrari discuss the oscillations as a
problem of resonant scattering of gravitational waves incident on a
potential barrier generated by the spacetime curvature.  The scattering
viewpoint is required for axial perturbations, which do not couple to
the star and have no newtonian counterpart; but even for polar
perturbations, a decoupling of matter and metric perturbations allows
one to view the perturbations as a scattering problem.  Although the
polar perturbations are reducible to a Schr\"odinger equation only
outside the star, Chandrasekhar and Ferrari find in the interior a set
of equations that involve only the perturbed metric.  Once the system
has been solved, the motion of the fluid can be obtained
algebraically.  A conserved current expresses a flux of gravitational
"energy" through a perturbed star in a way that generalizes the Regge
theory of resonant scattering to the present context.   The local
energy flux is gauge-dependent, but the total power radiated is not.

New classes of outgoing modes (which they call resonances) have
emerged from this work.  Outgoing axial modes can occur in usually
dense spherical models, while in rotating stars, the coupling of polar
and axial perturbations results in a family of rotationally induced
outgoing modes.

*

New models of rapidly rotating neutron stars have been constructed
recently by Cook, Shapiro and Teukolsky (Rapidly Rotating Neutron Stars
in General Relativity: Realistic Equations of State, Cornell preprint
CRSR 1047), and by Eriguchi, Hachisu and Nomoto (Proc. Roy. Astr. Soc.,
in press).  Both groups use codes based on a code developed initially
for relativistic polytropes by Komatsu, Eriguchi and Hachisu.  The Cook
et.  al. work presents sequences of stars with fixed baryon number,
which can be taken to represent the evolution of star that is losing
angular momentum.   Rotation supports stars with masses above the
maximum allowed for a spherical star, and Cook et. al. highlight a
feature of rotating sequences that was not noticed in previous models
considered by Friedman, Ipser and Parker.  As a neutron star loses
angular momentum, it ordinarily spins down, but near the lower limit on
angular momentum, the angular velocity increases as the angular
momentum decreases, and they suggest that the unexpected spin-up may
provide an observable precursor to collapse.

*

Ipser and Lindblom ({\it The Astrophysical Journal}, {\bf 389}, 392
(1992)) have begun the difficult work needed to generalize to
relativistic stars the elegant formalism they used in newtonian gravity
to compute normal modes of rotating stars.  For short wavelength modes,
one can use the "Cowling approximation," neglecting the change in the
gravitational potential.  The relativistic equations of motion then
have essentially the same character as the perturbed Euler equations,
and the hydrodynamic degrees of freedom of the adiabatic pulsations of
relativistic fluids can be described by a single scalar potential.
This potential is determined by a second-order (typically elliptic)
partial differential equation.  They obtain a  variational principle
from which the pulsation frequencies may be evaluated in this
approximation, and which can be generalized to an Eulerian principle
that does not require the Cowling approximation.

Work by Mendel and Lindblom (``Superfluid Effects on the Stability of
Rotating Newtonian Stars,'' in {\it The Structure and Evolution of
Neutron Stars}, ed. by D. Pines, R. Tamagaki and S. Tsuruta
(Addison-Wesley: 1992) pp.~224-226) on dissipation in rotating
superfluids shows an unexpectedly large internal dissipation in
rotating neutron stars.  The result apparently implies stablility
against nonaxisymmetric modes in neutron stars cooler than the
superfluid transition temperature of about $10^9$ K.  Fast pulsars with
weak enough magnetic fields would then be limited in their angular
velocity only by the Kepler frequency, the angular velocity of an
particle in orbit just outside the equator.

\vfill
\eject

\centerline{\bf Two Dimensional Black Holes}
\medskip
\centerline{Gary T. Horowitz,  UCSB}
\bigskip
\bigskip
There has been a great deal of interest in the past couple of years
on the subject of two dimensional black
holes.  One might well ask, ``Why is there so much interest in a
subject clearly removed from physical applications?". The answer is
two-fold.  Since two dimensions is clearly simpler than four, one can
use these black holes as toy models to study processes which cannot
yet be analyzed in higher dimensions. In addition, there are arguments
which show that certain properties of extremal four dimensional black
holes are accurately described by the two dimensional solution.

There are actually two quite different lines of research which are
being pursued. The first is to try to understand the endpoint of
Hawking evaporation of two dimensional black holes, and to answer the
question of whether pure states evolve to mixed states. An important
step in this direction was taken by Callan, Giddings, Harvey, and
Strominger (Phys. Rev. 45 (1992) R1005).  They study black holes in a
theory of a metric and scalar field (the dilaton) in two spacetime
dimensions. This theory can be quantized exactly using canonical
quantization, but it has no local degrees of freedom and hence no
Hawking radiation.  If one adds a scalar field, then it turns out that
one can find exact solutions describing the classical collapse of
matter to form a black hole. One can then calculate the Hawking
radiation in this classical background using the trace anomaly (as
originally pointed out many years ago by Christensen and Fulling).
One sees explicitly how the radiation ``turns on" and approaches a
constant rate at late times.

Perhaps the most important advantage of working in two dimensions is
that one can include the backreaction of the radiation on the black
hole.  However, even in two dimensions, it appears difficult to
compute all quantum effects exactly. If one considers a large number
$N$ of scalar fields, one can calculate an effective action to leading
order in $1/N$. Extrema of this effective action then represent the
evolution of a black hole including backreaction. The quantum
corrected equations are too complicated to solve exactly, but
numerical calculations show that the black hole evaporates in a finite
time. Unfortunately, when the black hole becomes very small, the $1/N$
approximation breaks down. In other words, the situation is similar to
the four dimensional case: One has to go beyond the $1/N$ or
semiclassical approximation to understand the endpoint of Hawking
evaporation.  However, it has been pointed out that there are many
inequivalent quantizations of this classical theory. In some of them
the quantum theory is exactly soluble. In their simplest form, the
soluble theories appear to be unphysical since the energy is unbounded
from below, and black holes continue to radiate indefinitely. However,
a modification has been proposed (corresponding to a boundary
condition at strong coupling) which cures this problem. It is
currently a matter of debate whether this modified theory preserves
quantum mechanical predictability.

The second line of research involving the black hole concerns its
connection to string theory. The equations of motion for the metric
and dilaton (which are solved by the 2D black hole) are part of the
low energy field equations for string theory. The exact string
equations include extra terms involving higher powers of the
curvature.  Witten showed how to construct an exact solution to string
theory which reduces to the black hole in the low energy limit (E.
Witten, Phys.  Rev. D 44 (1991) 314).  This was the first first time
that an exact black hole solution was found. (More recently, exact
black holes have also been found in three and four dimensions,
although these examples are not asymptotically flat.)  People are now
studying various aspects of this exact solution.  Perhaps the most
important is the existence of a singularity.  Singularities in string
theory should be defined by studying the motion of test strings. There
are now several examples of geodesically incomplete spacetimes,
including some with curvature singularities, for which the motion of
test strings is well behaved.  In other words, string theory
``resolves" certain curvature singularities classically, without
including quantum effects of spacetime.  (See, e.g. J. Horne, G.
Horowitz and A. Steif, Phys. Rev. Lett.  68 (1992) 568.)  The exact 2D
black hole spacetime has a curvature singularity, and the current
indications are that it {\it is} singular in string theory.  So the
quantum gravity effects will be important. Preliminary work is
underway to study the evaporation of the black hole in the full
context of string theory.

\vfill
\eject

\centerline{\bf Resonant-Mass Gravitational Wave Detectors - An Update}
\medskip
\centerline{Peter Michelson, Stanford}
\bigskip
\bigskip

Currently, two cryogenic resonant-mass detectors, operating at
4 K, are in continuous operation at Louisiana State University
and at CERN (University of Rome group).  The LSU detector has
been in nearly continuous operation for more than two years with
an rms sensitivity of $h = 6 \times 10^{-19}$.  They report less than one
non-Gaussian event per day.  Coincidence analysis of data from the
last half of 1991 was carried out with the Rome group and a new
upper limit on the g.w. flux impinging on the earth was reported
at the last international conference in Argentina.  This is an
improvement on the previous upper limit reported from a three-way
coincidence with detectors at LSU, CERN, and Stanford.

At Stanford, work is continuing on the assembly of a new ultralow
temperature detector designed to operate at 50 mK.  Fabrication of
the major components of the detector have been completed. Prof.
H.J. Paik's group at the University of Maryland is closely
collaborating with the Stanford group and are supplying a resonant-
mass superconducting inductive transducer for the Stanford detector.
The cryostat for the new system has successfully undergone a series
of leak tests from room temperature to 77 K.  It is expected that
the antenna will be installed in the cryostat by the end of this
year.

The Nautilus ultralow temperature detector of the Rome group has
been installed at the INFN laboratory in Frascati.  Frascati is
now the center of activity for the Rome group.  The INFN has
provided excellent new facilities for the gravity wave detection
effort there, including a new building.  An effort of similar
magnitude is underway at the INFN Laboratory in Legnaro, Italy.
The group there, headed by Massimo Cerdonio, recently sponsored
a Workshop on Resonant-Mass Detectors.  The meeting was attended
by about 100 people representing nearly every active group in the
field.  There was extensive discussion of currently operating
detectors, progress on ultralow temperature detectors, and on the
feasibility of constructing a new generation of massive, spherical
antennas.

In September, the US groups doing research on resonant-mass detectors
will hold a two day meeting to plan a detailed feasibility study
on spherical, resonant-mass detectors.  Issues about vibration
isolation and suspension, cryostat design, and antenna fabrication
will be addressed as well as formulation of a plan for coordinating
the efforts of the various groups involved in the study.

\vfill
\eject

\centerline{\bf Universality and Scaling in Gravitational Collapse}
\medskip
\centerline{Richard Matzner, University of Texas at Austin}
\bigskip
\bigskip

Recent results by Matthew Choptuik (Texas) on spherical scalar
collapse, and by Andrew Abrahams (Cornell) and Charles Evans (Chapel
Hill) on axisymmetric gravitational collapse, show a surprising
scaling behavior, and critical behavior similar to that found in many
phase transition phenomena.

The work by Choptuik is a high accuracy spherical study, with
minimally or nonminimally coupled scalar radiation interacting only
gravitationally.  He uses a mesh refinement scheme as suggested by
Berger and Oliger (1984), and is able to evaluate variations over $\sim
10^7$ in scale by adaptively refining the grid as appropriate.  The
Abrahams/Evans code uses a simpler type of adaptation which also
allows for increased resolution of key regions of the computational
domain when needed. It evaluates pure vacuum axisymmetric behavior,
imploding shells of gravitational radiation.

By using his adaptive techniques, Choptuik has demonstrated the
ability of the algorithm to resolve solution features on essentially
arbitrarily small spatial and temporal scales.  He has used it to
perform an exhaustive survey of the solution space of his model.
(Choptuik, 1992; 1993a) In the course of this survey a variety of
unanticipated, and apparently generic, features of strongly
self-gravitating pulses of massless scalar radiation were discovered.
They appear to be the manifestation of "critical" behavior in the
model: there are points in parameter-space where the details of the
solution exhibit an exponential dependence on the initial (t=0)
configuration of the scalar field.  The solution wave profile becomes
independent of the initial shape, and an echo is observed; repeated
transformed copies of a standard waveform emerge.  Further, a generic
parameter dependence for the formation of black holes is found.
Choptuik sets up a sequence of examples of localized spherical
imploding waves which differ only in one parameter, say wave
amplitude.  As the amplitude of the wave approaches the critical value
which initiates black hole formation the final outgoing scalar wave
profile takes on the characteristic behavior.  A series of waveforms
is emitted which are of roughly constant amplitude, but of
logarithmically self similar shape in space and time.  As a particular
evolution approaches a limit time, the wave pattern repeats with a
time compression of about $e^{3.4}$ per repeat; there is also a
logarithmic self similar shape in space.  Choptuik can estimate from
the numerics both the time and spatial compression factor and finds
them closely equal.  Because of the extremely fine zoning allowed by
his local grid refinement technique, Choptuik is able to determine the
value of the parameter (here the amplitude) leading to black hole
formation, to machine precision ($\sim 1$ in $10^{13}$ on a Cray
machine).  As the amplitude is adjusted to lie just above the critical
value a*, a black hole forms, and a remarkable analog of other
critical phenomena appears: $m_{BH} = m_f |a-a*|^\gamma$ where $m_f$ is
a constant and $\gamma$ is an exponent numerically determined by
Choptuik: $\gamma = 0.37...$ .  This means that black holes of
arbitrarily small mass can form.  As the amplitude approaches a* from
above, one also sees a logarithmic repetition of outgoing waveforms.
What is truly remarkable and particularly characteristic of critical
phenomena is that the same scaling holds regardless of the control
parameter; for instance, narrowness of the pulse can be used to
control the formation or not of the black hole; the same echoing and
the same critical exponent are found.  Further, this behavior with the
same exponent holds for black hole formation from infalling minimally
coupled scalar fields (Choptuik 1992, 1993a), and nonminimally coupled
scalar fields (Choptuik 1992, 1993b).  The extreme dynamic range of
Choptuik's results $\sim 10^7$ is such that, even with today's
fastest machines, they would have been extremely difficult to obtain
without his use of the adaptive algorithm.

Remarkably similar results were obtained by Abrahams and Evans, for
gravitational wave implosions to form black holes.  (Abrahams 1992,
Evans 1992, Abrahams and Evans 1993) Their code is computationally
complicated because it is axisymmetric rather than spherical.
Although the mesh structure is fixed, grid adaptation is achieved via
a grid velocity which provides increased central resolution during the
period of strong field dynamics and/or black hole formation.  The
Abrahams/Evans code explores a very important direction in the physics
of this phenomenon since it is nonspherical and vacuum.  Evans and
Abrahams find very close agreement for the black hole mass behavior,
finding the same dependence of the black hole mass on the wave
strength, with the same power gamma.  They also find an echoing
similar to that in Choptuik's scalar case, but with substantially
different exponent.  Whereas Choptuik's rescaling occurs with a factor
$e^{3.4} \sim 30$, Abrahams and Evans find apparently $e^{0.6} \sim
1.8$ for the factor.  Better nonspherical collapse codes may
eventually be needed to verify the echoing result, but the black hole
mass result seems very convincing.  The exponent 0.37 for the formed
mass of the black hole, incidentally, is solidly in the range of
critical exponents for condensed matter phase transitions.

\parindent=0pt
\parskip=5pt

A. Abrahams, in Syracuse Workshop on Numerical Relativity/Black Holes,
G. Fox, editor (1992).

A.M. Abrahams and C.R. Evans, "Critical Behavior and Scaling in Vacuum
Axisymmetric Gravitational Collapse," Phys. Rev. Lett., 70, p. 2980 (1993).

M.S. Berger and J. Oliger, J. Comp. Phys. 53 p. 484 (1984).

M. Choptuik, "Critical Behavior in Massless Scalar Field Collapse,"
in Approaches to Numerical Relativity, R. d'Inverno, ed., Cambridge
University Press, (1992).

M. Choptuik, "Universality and Scaling in Gravitational Collapse of a
Massless Scalar Field," Phys. Rev. Letters, 70, p. 9 (1993a).

M. Choptuik, private communication (1993b).

C.R. Evans, in Syracuse Workshop on Numerical Relativity/Black Holes, G.
Fox, editor (1992).
\parskip=\baselineskip
\vfill
\eject
\parindent=10pt

\centerline{\bf Gravitational Wave Memories Upgraded:
The Christodoulou Effect}
\medskip
\centerline{Richard Price, University of Utah}
\bigskip
\bigskip
The physical effect of a gravitational wave is the displacement of
free masses by the strain $h_{ij}$ of the wave. If a burst of waves
pass, the stationary value of $h_{ij}$ after the passage of the waves
will not be the same as that before the waves. In principle this would
leave the free masses of a gravitational wave detector in positions
different than their positions before the waves arrived, and the
shifted positions would constitue a souvenir of the burst. That
gravitational (and other) wave trains have such ``memory" has long
been known, as has the connection with the ``zero frequency limit"
(ZFL) of the fourier decomposition of the wave amplitude. In practice
this ZFL feature is the basis for detection. The waves in the burst
need only to be measured at a frequency below the lowest natural
frequency in the wave burst (i.e., the reciprocal of the burst
duration) [1]. The measurement of the true DC offset of detector
masses would be swamped by drift and by low frequency noise.

The reason for recent interest in wave memories is the correction of a
widespread and longstanding conceptual error concerning their
calculation.  It had been accepted that the calculation of the memory
produced by an astrophysical event required only linearized gravity
theory, even for a strong field events like black hole collisions. The
argument went something like this: Only $1/r$ contributions to
$h_{ij}$ are radiative; for the memory only the initial and final
values of $h_{ij}$ are needed; before and after the strong
accelerations producing the fat middle of the burst, the stress energy
is simply that of ``particles" moving at uniform velocity; the $1/r$
fields of these particles is just the Coulomb fields due to their
masses, and those contributions are completely within the scope of
linear theory.

Since the memory of a nonlinear interaction could be calculated with
linear theory, it seemed to give a cheap partial answer to questions,
e.g, about the radiation from a collision of black holes.[2] There
were some disquieting unresolved issues. Philip Payne [3], for
instance, noticed that the ZFL argument for black holes led to
inconsistencies. A nonlinear contribution to the memory was present in
the studies of equation of motion by Luc Blanchet and Thibault Damour
[4], but its importance escaped notice.  All this changed about a year
ago when Demetrious Christodoulou, with a careful and mathematically
rigorous analysis, showed that the memory does in general have
nonlinear contributions [5]. This nonlinear contribution is now called
the ``Christodoulou part of the memory."

The flaw in the linear-is-adequate analysis is that in the final
state, after the passage of the wave burst, the sources of the field
are not only the particles in uniform motion; there is also the
gravitational radiation of the burst. This turns out to be
quantitatively, as well as pictorially, the origin of the
Christodoulou contribution. As Kip Thorne has pointed out, the
previously believed relationship between memory and the asymptotic
states of the interacting particles remains valid if the gravitons of
the final state are included[6]. Since the stress-energy of the
gravitons generated in a strong field interaction (e.g., black hole
collision) can be comparable to that for the initial and final states
of the holes, the graviton contribution to the memory cannot be
ignored.

This picture helps clarify aspects of the Christodoulou contribution,
such as why there is interest in gravitons generating memory (i.e.,
zero frequency radiation) but not in their generation of waves at
other frequencies. The answer lies in the fact that the gravitational
wave stress- energy originates in the motion of the stress-energy of
the source.  Each orbit of, e.g., a binary neutron star will generate
an amount of wave energy rather small in comparison with the particle
(i.e., neutron star) mass-energy.  For each orbit, then, the radiation
is a considerably weaker source of waves than the stars themselves.
For the memory, however, it is the totality of the energy generated
over several or many cycles that competes in strength with the ``real"
stress-energy of the stars.

Calculations by Alan Wiseman and Clifford Will[7] are in accord with
this picture. They considered gravitational bremsstrahlung, the small
angle scattering of one mass (hole, star, etc) by another, and find
that the Christodoulou contribution to memory must be small. (There is
relatively little radiation generated, and only one ``cycle.") For
coalescing neutron star binaries, the presently favored source for
wave bursts, the situation is quite different. The complete waveforms
for the coalescence are not known, but Wiseman and Will use a model
calculation to estimate that the Christodoulou contribution to the
memory may be as much as 20\% of the amplitude of the quadrupole
radiation.

It has been a bit of an embarrassment for a widely accepted ``truth"
to be wrong. Kip Thorne has claimed, in print[6], that this defeat of
handwaving by mathematics has been a ``salubrious experience."
Presumably we will all wave our hands a bit more carefully for a
while, but the experience might turn out to be useful for more than
just building character. The Christodoulou contribution gives a
measure of the total wave power generated. For binary neutron star
coalescence the total power will depend on the details of the final
tidal disruptions, in particular on the radii of the constituent
neutron stars. The Christodoulou contribution to the memory of the
wavetrain might give us the most precise observable measure of this
neutron star physics.[8]

\parindent=0pt
\parskip=3pt

[1] For a discussion of the measurement of wave memory, see V. B.
Braginsky and K. S. Thorne, Nature (London) 327, 123 (1987).

[2] L. L. Smarr, Phys. Rev. D 15, 2069 (1977).

[3] P. N. Payne, Phys. Rev. D 28, 1894 (1983).

[4] L. Blanchet and T. Damour, in Proceddings of the 12th
International Conference on General Relativity and Gravitation, 1989,
Conference Abstracts (unpublished), p. 265. L. Blanchet, Habilitation
thesis, Universite Pierre et Marie Curie, Paris, 1990, pp. 205-207.

[5] D. Christodoulou, Phys. Rev. Lett. 67, 1486 (1991).

[6] K. S. Thorne, Phys. Rev. D 45, 520 (1992).

[7] A. G. Wiseman and C. M. Will, Phys. Rev. D 44, R2945(1991).

[8] E. Flanagan, D. Kennefick, D. Markovic, and K. S. Thorne, paper in
preparation.
\parindent=10pt
\parskip=\baselineskip

\vfill
\eject

\noindent{\it Conference Report: }

\centerline{\bf Topical Conference on Quantum Aspects of Balck Holes}
\centerline{\it Santa Barbara June 21-27 1993}
\medskip
\centerline{ Jeff Harvey, Gary Horowitz, and  Andy Strominger, Coordinators}
\bigskip
\bigskip

The topical conference on Quantum Aspects of Black Holes was held from
June 21-27 1993 and was attended by 113 physicists, including most of
the world's experts on the quantum physics of black holes.  The
conference followed two separate but related workshops held at the
ITP, one entitled {\it Non-Perturbative String Theory}, and the other
{\it The Small-Scale Structure of Space-Time}.

Different aspects of black holes, string theory, and quantum gravity
were discussed with the focus on the question of whether the physics
of black holes requires a fundamental change in the laws of quantum
mechanics.  The possibility of an irreconcilable difference between
black hole physics and quantum predictability was raised in 1975 by
S.~Hawking and emphasized in his presentation at the conference.
Hawking argued that black holes decay by emitting a thermal spectrum
of particles, thus apparently losing any information about the matter
which originally collapsed to form the black hole.  This loss of
information contradicts quantum mechanical predictability which
requires microscopic preservation of information.

Interest in this question has been revived in the last two years by
the discovery of simplified models of black holes which allow a more
precise treatment of quantum corrections to the process of Hawking
evaporation.  At the time of the conference those in attendance were
split into two definite camps: those believing in loss of information,
and those arguing that quantum predictability need not be given up.
One of the outcomes of this conference was a fairly detailed set of
problems that each camp must address.  Roughly speaking, the
information loss camp needs to develop a specific set of calculational
rules which supplant the usual deterministic rules of quantum
mechanics and does not allow unacceptably large violations of energy
conservation or causality at low energies.  The quantum predictability
camp has had difficulty in providing a plausible explanation of how
the information is retained in the radiation coming from the black
hole and has had to postulate the existence of new microscopic degrees
of freedom at the horizon of the black hole.  Their task is to provide
a specific model of these new degrees of freedom and at least a
heuristic description of how they can restore information to the
Hawking radiation.

The conference did not resolve this fundamental question, but it did provide
a framework for future results, sharpened the problems which need to be
addressed, and left participants with new energy and enthusiasm for
tackling these problems.

\eject
\noindent{\it Conference Report:}

\centerline{\bf Workshop on knots and Quantum Gravity}
\centerline{\it Riverside May 14-16 1993}
\medskip
\centerline{John Baez, University of California at Riverside}
\bigskip
\bigskip

The loop representation is a promising approach to the quantization of
gravity, in that it is nonperturbative and manifestly respects the
general covariance of the theory.  The physics behind it is
conservative, essentially amounting to the canonical quantization of
Einstein's equations.  It is, however, mathematically innovative, and
has forged connections between general relativity and the
superficially unrelated subject of knot theory.  The goal of this
workshop was to bring together physicists and topologists and begin a
dialog that might catalyze further research on these connections.  It
took place on May 14th-16th at the University of California,
Riverside, under the auspices of the mathematics and physics
departments.  The proceedings will appear in a volume entitled Knots
and Quantum Gravity, to be published by Oxford University Press.

On Friday the 14th, Dana Fine spoke on ``Chern-Simons theory and the
Wess-Zumino-Witten model.''  Witten's original work deriving the Jones
invariant of links (or, more precisely, the Kauffman bracket) from
Chern-Simons theory used conformal field theory as a key tool, and by
now the relationship between conformal field theory in 2 dimensions
and topological quantum field theories in 3 dimensions has been
explored from a number of viewpoints.  The path integral approach,
however, has not yet been worked out in full mathematical rigor.  In
this talk Fine described work in progress on reducing the Chern-Simons
path integral on $S^3$ to the path integral for the Wess-Zumino-Witten
model.

Also, Oleg Viro spoke on ``Simplicial topological quantum field
theories."  Recently there has been increasing interest in formulating
topological quantum field theories (TQFTs) in a manner that relies
upon triangulating spacetime.  In a sense this is an old idea, going
back to the Regge-Ponzano model of Euclidean quantum gravity.
However, this idea was given new life by Turaev and Viro, who
rigorously constructed the Regge-Ponzano model of 3d quantum gravity
as a TQFT based on the $6j$ symbols for the quantum group $SU_q(2)$.
Viro discussed a variety of approaches of presenting manifolds as
simplicial complexes, cell complexes, etc., and methods for
constructing TQFTs in terms of these data.

Saturday's talks began with a thorough review of recent work on the
loop representation of quantum gravity by Renate Loll, Abhay Ashtekar
and Jorge Pullin.  Loll's talk was entitled ``Loop formulation of
gauge theory and gravity.''  This served to introduce the loop
representation and its various physical applications.  Ashtekar's
talk, ``Loop transforms,'' largely concerned his new work with Jerzy
Lewandowski on making the loop representation into rigorous
mathematics.  The notion of measures on the space ${\cal A}/{\cal G}$
of connections modulo gauge transformations has long been a key
concept in gauge theory, which however has been notoriously difficult
to make precise.  The key notion developed by Ashtekar and Isham for
this purpose is the ``holonomy C*-algebra,'' an algebra of observables
generated by Wilson loops.  Formalizing the notion of a measure on
${\cal A}/{\cal G}$ as a state on the holonomy C*-algebra, Ashtekar
and Lewandowski have been able to construct such a state with
remarkable symmetry properties, in some sense the natural analog of
Haar measure on a compact Lie group.  Ashtekar also discussed the
implications of this work for the study of quantum gravity.

Pullin spoke on ``The quantum Einstein equations and the Jones
polynomial.''  Perhaps the most remarkable connection between knot
theory and quantum gravity is that, using the loop representation, the
Jones polynomial invariant of links (or more precisely, the Kauffman
bracket) represents a state of quantum gravity with cosmological
constant, essentially a quantization of anti-deSitter space.  Pullin
detailed his work with Bernd Br\"uegmann and Rodolfo Gambini on this
subject.  He also sketched the proof of a new result, that the
coefficient of the 2nd term of the Alexander-Conway polynomial
represents a state of quantum gravity with zero cosmological constant.

On Saturday afternoon, Louis Kauffman spoke on ``Vassiliev invariants
and the loop states in quantum gravity.''  One aspect of Br\"uegmann,
Gambini and Pullin's work that is especially of interest to knot
theorists is that it involves extending the bracket invariant to
generalized links admitting certain kinds of self-intersections.  This
concept also plays a major role in the study of Vassiliev invariants
of knots.  Curiously, however, the extensions occuring in the two
cases are different.  Kauffman explained the relationship between the
two from the path-integral viewpoint.

Sunday morning began with a talk by Gerald Johnson, ``Introduction to the
Feynman integral and Feynman's operational calculus,'' on his work
with Michel Lapidus on rigorous path integral methods.  Viktor
Ginzburg then spoke on ``Vassiliev invariants of knots,'' and in the
afternoon, Paolo Cotta-Ramusino spoke on ``4d quantum gravity and knot
theory.''  This talk dealt with his work in progress with Maurizio
Martellini.  Just as Chern-Simons theory gives a great deal of
information on knots in 3 dimensions, there appears to be a
relationship between a certain class of 4-dimensional theories and
so-called ``2-knots,'' that is, embedded surfaces in 4 dimensions.
This class includes quantum gravity in the Ashtekar formulation,
Donaldson theory, and the so-called $B \wedge F$ theory.
Cotta-Ramusino and Martellini have described a way to construct
observables associated to 2-knots, and are endeavoring to prove at a
perturbative level that they give 2-knot invariants.

Louis Crane spoke on ``Quantum gravity, spin geometry, and categorical
physics.''  This was a review of Crane's work on the use of
2-categories in 4-dimensional quantum gravity and his construction
with David Yetter of a 4-dimensional TQFT based on the ``15j symbols''
for $SU_q(2)$.  Just as every 3d TQFT gives rise to a braided tensor
2-category, and thus solutions to the Yang-Baxter equations, Crane's
generalization of TQFT suitable for 4 dimensions gives rise to a
braided tensor 2-category, and thus solutions to the Zamolodchikov
tetrahedron equations.  In the conference proceedings there will also
appear a paper by J.\ Scott Carter and Masahico Saito, ``Knotted
Surfaces, Braid Movies, and Beyond,'' dealing with their work on
2-knots, 2-braids and 2-tangles.   All these topological objects can
be drawn as movies in which links evolve in time, with one
``elementary string interaction'' occuring between each frame.   Just
as any two pictures of the same knot can be bridged by a sequence of
Reidemeister moves, Carter and Saito have developed a set of ``movie
moves'' relating any two movies of the same 2-knot.  More recently,
they have developed a theory of 2-braids.  Their paper discusses these
matters and also reviews new solutions of the Zamolodchikov
tetrahedron equations.

In the final talk of the workshop, John Baez spoke on ``Strings,
loops, knots and gauge fields.''  He attempted to clarify the
similarity between the loop representation of a generally covariant
gauge theory and a topological string theory.  At a fixed time both
involve loops or knots in space, but the string-theoretic approach
should clarify the role of surfaces embedded in spacetime.

\vfill
\eject
\noindent{\it Conference Report: }

\centerline{\bf NATO Advanced Research
Workshop on Deterministic Chaos }
\centerline{\bf in General Relativity}
\centerline{\it
Kananaskis, Alberta, Canada,26--30 July 1993}
\medskip
\centerline{Beverly Berger, Oakland University}
\bigskip
\bigskip

The Workshop was organized to address a controversy that had arisen in
recent years over whether or not the dynamics of the vacuum, diagonal
Bianchi IX (Mixmaster) cosmology is chaotic.  Although most of the
discussion concerned this topic, the focus of the meeting broadened to
include other exciting nonlinear phenomena within general relativity.

The workshop began with an overview of the main issues by Hobill.  It
has long been known that the approach to the singularity of the
Mixmaster model can be approximated as an infinite sequence of Kasner
epochs.  The epoch changes when the trajectory in minisuperspace (MSS)
bounces off the MSS potential (spatial scalar curvature).  In the
anisotropy plain, the trajectory bounces between two walls which form
a corner of the MSS potential, moving outward from the origin
(isotropy) but with the trajectory angle with the outward pointing
corner ray increasing after each bounce.  At some point, the
trajectory points inward rather than outward so that it moves to a new
corner where the process begins again.  This represents the start of a
new era.  In the late 1960's, Belinskii, Lifshitz, and Khalatnikov
(BKL) showed that this evolution could be expressed as a discrete map
for $u$ which parametrizes each Kasner epoch.  Within a single era,
the map is $u_{n+1}=u_n-1$ (for $u_n \ge 2$).  At the end of an era,
$u_{n+1} = \left( u_n-1 \right)^{-1}$ (for $1 \le u_n \le 2$).  It is
only the era change which is sensitive to initial conditions (SIC).
The sequence of era changes is the chaotic Gauss map $u_{N+1}= \left(
u_N - [u_N] \right)^{-1}$ (where [ ] denotes integer part and $u_N$ is
the $u$ value that starts each era).  One measure of chaos is the
Liapunov exponent (LE) which measures the divergence of initially
nearby trajectories and is positive for chaotic systems.  (The Gauss
map has a LE $> 0$.  In the late 1980's, it became possible to
evaluate LE's for numerical solutions of ordinary differential
equations.  Application of these techniques to Mixmaster by
independent groups (e.g. Burd et al, Hobill et al) yielded the
surprising result of vanishing LE implying that the Mixmaster dynamics
is not chaotic in direct conflict with the Gauss map result.  To
further confuse the issue, the Mixmaster LE values depended on the
choice of variables used, leading to LE $> 0$ in some cases (e.g.
Pullin).  Various analyses purporting to demonstrate Mixmaster to be
either chaotic or not chaotic then began to appear in the literature.
It became clear that the difficulty with the LE is due to the
exponential increase in epoch duration in terms of some of the time
coordinates used.  However, there remained significant confusion over
the exact nature of Mixmaster dynamics and whether or not an
appropriate measure could be found to answer the chaos question once
and for all.

To provide a framework for the discussion of these questions at the
workshop, introductions to chaotic dynamics were given by Churchill
and Bishop while MacCallum provided an overview of relativistic
cosmologies.  Misner reviewed both his original formulation and that
of the Misner-Chitre coordinates which allow an approximate treatment
of the model as geodesic flow on a space of negative curvature with a
MSS potential of fixed location in the relevant anisotropy plane.
Coley presented Wainwright's variables (based on the analysis of
Bogoyavlensky) which allow unification of the treatment of all Class A
Bianchi Types and can be used to show rigorously how Bianchi Types I
and II attract the Bianchi IX solution between bounces and during the
bounce respectively.  Berger used an amalgam of the BKL and Misner
approaches to demonstrate the relationship of the BKL approximate
discrete evolution to the Mixmaster dynamics.  Additional observations
on all these approaches were given by Rugh, Burd, and Creighton.  The
juxtaposition of this variety of treatments demonstrated that
Mixmaster dynamics itself is well-understood with SIC at an era change
only.  This ``weak'' SIC must be reflected correctly for any measure
of chaos in the model to be regarded as successful.  Yet the issue of
chaos itself may be semantic rather than substantive since the answer
given by any measure may depend on the weighting it gives to the
regular and SIC parts of the Mixmaster dynamics.

Other solutions in general relativity exhibit interesting nonlinear
dynamics.  Choptuik discussed the strange echoing wave form and
critical scaling he has observed in a numerical study of the collapse
of a spherically symmetric scalar field.  Berger displayed the
competition between the growth of small scale structure and the
development of velocity dominance seen numerically in the approach to
the singularity of the unpolarized Gowdy $T^3$ cosmology.  Chrusciel
applied dynamical systems theory to the properties of Cauchy horizons
required by Hawking's Chronology Protection Conjecture.  Calzetta
discussed the application of a method using homoclinic orbits to test
for chaos in both test body trajectories in perturbed black hole
spacetimes and in a Friedmann-Robertson-Walker (FRW) model with
conformally coupled scalar field.  Tavakol used the chaotic
trajectories possible in closed negatively curved FRW models (with
nontrivial topology) to study the anisotropy of the cosmic microwave
background.  Ribeiro introduced a fractal cosmology in order to
reproduce the observed galaxy-galaxy correlation function.  Tomaschitz
discussed chaos in the context of quantum fields on curved spacetime.

The 25 participants found the discussion free-ranging and stimulating.
The interactions between mathematicians and physicists, between
dynamicists and relativists made the workshop particularly worthwhile.
The proceedings, edited by Hobill et al, will be published by Plenum.

\bye